\documentclass[aps,pre,a4,twocolumn,groupedaddress,showpacs,floatfix]{revtex4}
\usepackage{amsmath}
\usepackage{amssymb}
\usepackage{graphicx}
\usepackage{epic}
\usepackage{eepic}
\usepackage{subfigure}
\begin{document}

\title{Morphological relationship between axon and dendritic
arborizations as revealed by Minkowski functionals}
\author{Marconi Soares Barbosa$^1$ and Luciano da Fontoura Costa$^1$}
\affiliation{$^1$Institute of Physics at S\~ao Carlos,
University of S\~ ao Paulo, S\~{a}o Carlos,
SP, PO Box 369, 13560-970,
Phone +55 16 3373 9858,
FAX +55 162 71 3616,
Brazil,
marconi@if.sc.usp.br}
\date{\today}
\begin{abstract}
The spatial structure of the axonal and dendritic arborizations is
closely related to the functionality of specific neurons or neuronal
subsystems.  The present work describes how multiscale Minkowski
functionals can be used in order to characterize and compare the
spatial organization of these two types of arborizations.  The
discrimination potential of the method is illustrated with respect
to three classes of cortical neurons.
\end{abstract}


\maketitle

Human intelligence is to a great extent determined by the
connectivity between the myriad of neuronal cells in the nervous
system (Cajal:1989).  Biologically, these connections are
implemented by the \emph{synapses} between neurons, which are
strongly related to the shape of axons and dendrites
(e.g.~\cite{Costa_can:2005, pelt:2002b, Ascoli:2004}).  For
instance, neural cells exhibiting more intricate dendritic or axonal
arborizations tend to promote more connections. Therefore, the
systematic study of the morphological properties of the axonal and
dendritic trees can provide essential information about the
connectivity of the nervous system.  Because neurite outgrowth can
be understood as a complex dynamical system involving pattern
formation, biochemical regulation and electrical activity, they are
particularly suitable for characterization and modeling by using
concepts and methods from physics, computation and biology.

As the spatial distribution of synapses is to a great extent
influenced by the interaction between the shape of dendrites and
axons, and also because they share several developmental aspects, one
question of particular relevance concerns the study to which extent
these two types of structures are related (i.e. similar or distinct).
Interestingly, both axons and dendrites start their lives as
\emph{neurites}, being differentiated during development.  Along the
early stages of neuronal growth, which involve many outgrowths and
retractions, one of these neurites dominates and become the cell axon,
assuming distinct biomolecular composition and function.  In
particular, axons seek actively for targets while being influenced by
many trophic factors including differential adhesiveness,
galvanotropism and chemotropism (e.g.~\cite{Sanes:2005, Cline:2001,
Kiddie:2005}).  Consequently, the shape of axons is directly affected
by external field influences.  At the same time, both dendrites and
axons share intrinsic genetic and biochemical basis, which constrains
neuronal growth and may imply intrinsic similarities between the
dendritic and axonal arborizations of a same cell or category of
cells.  The current article addresses this important problem from the
objective perspective of neuronal shape quantification by additive
functionals~\cite{Raedt:2001}, which have been previously applied with
encouraging success to the characterization of dendritic morphology
(e.g.~\cite{Barbosa:2003a,Barbosa:2003b}).

Integral geometry algorithms have been successfully used to
characterize morphologically complex patterns and structures whose
process of formation is not precisely known and is subject of
modeling~\cite{Raedt:2001}. The central procedure is the calculation
of intrinsic volumes or Minkowski functionals, a generalization of
the usual determination of volume.

The Minkowski functionals of a body $K$ in the plane are
proportional to familiar geometric quantities, namely its area
$A(K)$, perimeter $U(K)$ and the connectivity or Euler number
$\chi(K)$.  The usual definition of the connectivity from algebraic
topology in two dimensions is the difference between the number of
connected $n_c $ components and the number of holes $n_h$,
$\chi(K)=n_c-n_h$. In the Euclidean space, there is an additional
geometric quantity, the mean curvature or breadth. Moreover there
are two kinds of holes to consider: A pure hole, which is a
completely closed region of white voxels surrounded by black pixels
and tunnels. The Euler characteristic for the Euclidean space is
then given by
\begin{equation}
\chi(K)= n_c -n_t +n_h,
\end{equation}
where $n_t$ is the number of tunnels and $n_h$ is the number of pure
holes. One instance where these functionals appear naturally is
while attempting to describe the change in volume as the body $K$,
now assumed to be convex, undergoes a dilation through a parallel
set process using a ball $B_r$ of radius r
\begin{equation}
\label{eq:dila} V(K \oplus B_r)=V(K)+S(K)r+2\pi B(K) r^2
+\frac{4\pi}{3}r^3.
\end{equation}  Generalizing to higher dimensions, the change in hyper volume is given
by the Steiner formula
\begin{equation}
\label{eq_dilageral}
v^d(K\oplus B_r)=\sum_{\nu=0}^d \binom{d}{\nu}
W_{\nu}^{(d)}(K)r^{\nu},
\end{equation}
where the coefficients $W_{\nu}^{(d)}$ are referred to as Minkowski
functionals. For instance in the Euclidean space ($d=3$),
\begin{align}
\label{eq:familiar}
&W_0^{(3)}(K)=V(K),\quad W_1^{(3)}(K)=\frac{S(K)}{3},\\ \nonumber
&W_2^{(3)}(K)=\frac{2\pi}{3}B(K),\quad W_3^{(3)}(K)=\frac{4\pi}{3}\chi(K).
\end{align}
Despite the wealth of results and continuum formulae for obtaining
these functionals, it is useful to resource to the discrete nature of
the binary images we wish to analyze by looking at the distribution
of voxels in a Euclidean spatial lattice. By exploring the
additivity of the Minkowski functionals their estimation resumes
to counting the multiplicity of basic building blocks that
disjointedly compose the object. The fundamental information needed
here is a relationship for the functionals of an open interior of a
$n$-dimensional body $K$ which is embedded into a $d$-dimensional
space
\begin{equation}\label{eq_open}
W_{\nu}^{(d)}(\breve{K})=(-1)^{d+n+\nu}
W_{\nu}^{(d)}(K), \nu=0,\ldots, d.
\end{equation}
With the absence of overlap between these building blocks and using
the property of additivity of these functionals, we may write for a
pattern $\mathcal{P}$ composed of disjoint convex interior pieces
$\breve{N}_m$,
\begin{equation}\label{eq_whole}
W_{\nu}^{(d)}(\mathcal{P})=\sum_{m}
W_{\nu}^{(d)}(\breve{N}_m)n_m(\mathcal{P}), \; \nu=0,\ldots,d\quad,
\end{equation}
where $n_m(\mathcal{P})$ stands for the number of building elements
of each type $m$ occurring in the pattern $\mathcal{P}$. For the
three-dimensional space we display in Table~\ref{table:build3d} the
value of Minkowski functionals for the building elements in a
orthogonal lattice of voxels. By using the information (with $a=1$)
presented in Table~\ref{table:build3d} and Equations
\eqref{eq:familiar}, \eqref{eq_open} and \eqref{eq_whole} we have
\begin{align}
&V=n_3, \quad S=-6n_3+2n_2, \\ \nonumber
&2B=3n_3-2n_2+n_1,\quad \chi=-n_3 +n_2 -n_1+n_0,
\end{align}
Where $n_3$ is the number of interior cubes, $n_2$ is the number of
open faces, $n_1$ is the number of sides and $n_0$ is the number of
vertices. So the procedure to calculate Minkowski functionals of a
pattern $\mathcal{P}$ has been reduced to the proper counting of the
number of elementary bodies of each type that compose a voxel
(cubes, faces, edges and vertexes) involved in the make up of
$\mathcal{P}$.
\begin{table}[htb]
\begin{center}
\begin{tabular}{llllll}\hline\hline
$m$&  $\breve{N}_m$ &{$W_0^{(3)}$}&{$W_1^{(3)}$}&{$W_2^{(3)}$}&{$W_3^{(3)}$} \\ \hline
0 &  $\breve{V}$   & 0           & 0           & 0           &  1 \\ \hline
1 &  $\breve{L}$   & 0           & 0           & 1/2         &  1 \\ \hline
2 &  $\breve{F}$   & 0           & 2           & 1           &  1 \\ \hline
3 &  $\breve{C}$   & 1           & 6           & 3/2         &  1 \\ \hline
\hline
\end{tabular}
\caption{Minkowski functionals $\mathcal{M}_{\nu}^{(3)}$ for open elementary
 open sets $\breve{N}_m$ which compose a voxel $K$(of side
 $a=1$): $\breve{V}$ ({\it vertex}), $\breve{L}$ ({\it side}),  $\breve{F}$
 ({\it face}) e $\breve{C}$ ({\it cube}). \label{table:build3d}}
\end{center}
\end{table}

The recent effort aimed at replicating a working micro column of
cortex tissue~\cite{Markram:2006} has motivated the dissemination of
comprehensive data bases containing different anatomical classes of
neuronal cells. We have considered several neurolucida data from
anatomically different types of cells, namely the anatomical classes
MC1, BTC and LBC that populate the cortex and are engaged in
distinct functions in the working of a neocortical micro column. The
morphological data was obtained from the Brain Mind Institute, EPFL
database~\cite{database:2003}. Exact dilations of both the axonal
and dendritic trees were performed until the majority of holes and
tunnels disappeared, i.e., when the Euler number approaches 1. This
implied a total number of 1000 dilations.  All four Minkowski
functionals were calculated along such dilations, so as to obtain a
morphological signature for the dendritic and axonal arborizations
of each cell. We then conducted a morphological analysis using a
subset those values at selected scales, namely after 1,250, 500, 750
and 1000 exact dilation steps.  In order to reduce the effect of the
size of the cells on the measurements, the functionals were
normalized respectively to the adequate power of the diameter of the
cell, i.e the volume functional was normalized by $L^3$, the surface
area was normalized by $L^2$ and the mean-curvature was normalized
by $L$. The Euler-number is a dimensionless topological measure and
consequently did not undergo normalization. All measures were
subsequently standardized, which is accomplished by subtracting the
average and dividing by the standard deviation of each respective
type of measurement (i.e. one of the 4 functionals).  The
standardized values therefore have zero means and unit variance, and
most of their values are comprised between -2 and 2.  Such a
standardization reduces the influence of the overall relative
magnitude of the different measurements.

The functionals were calculated separately for the dendritic and
axonal trees of each neuron.  For each case, the obtained
measurements were organized into morphological vectors in the
20-dimensional space defined by the five spatial scales of each of
the four functionals. Therefore, given a neuronal cell represented
by their respective measurement vector, the morphological difference
between its axon and dendritic arborizations can be quantified in
terms of the magnitude $\mu$ of the vector corresponding to the
difference between the respective feature vectors (i.e. axon and
dendrites). Another important descriptor of the morphological space
which has been considered in this work is the angle $\phi$ between
the above difference vector and the vector which represents the mean
of all such difference vectors.  Such a measurement provides and
indication of how much the axon-dendrite in each cell departs from
the overall prototypical case (i.e. the average).

\begin{figure*}[htb]
\begin{centering}
 \includegraphics*[scale=.3,angle=0]{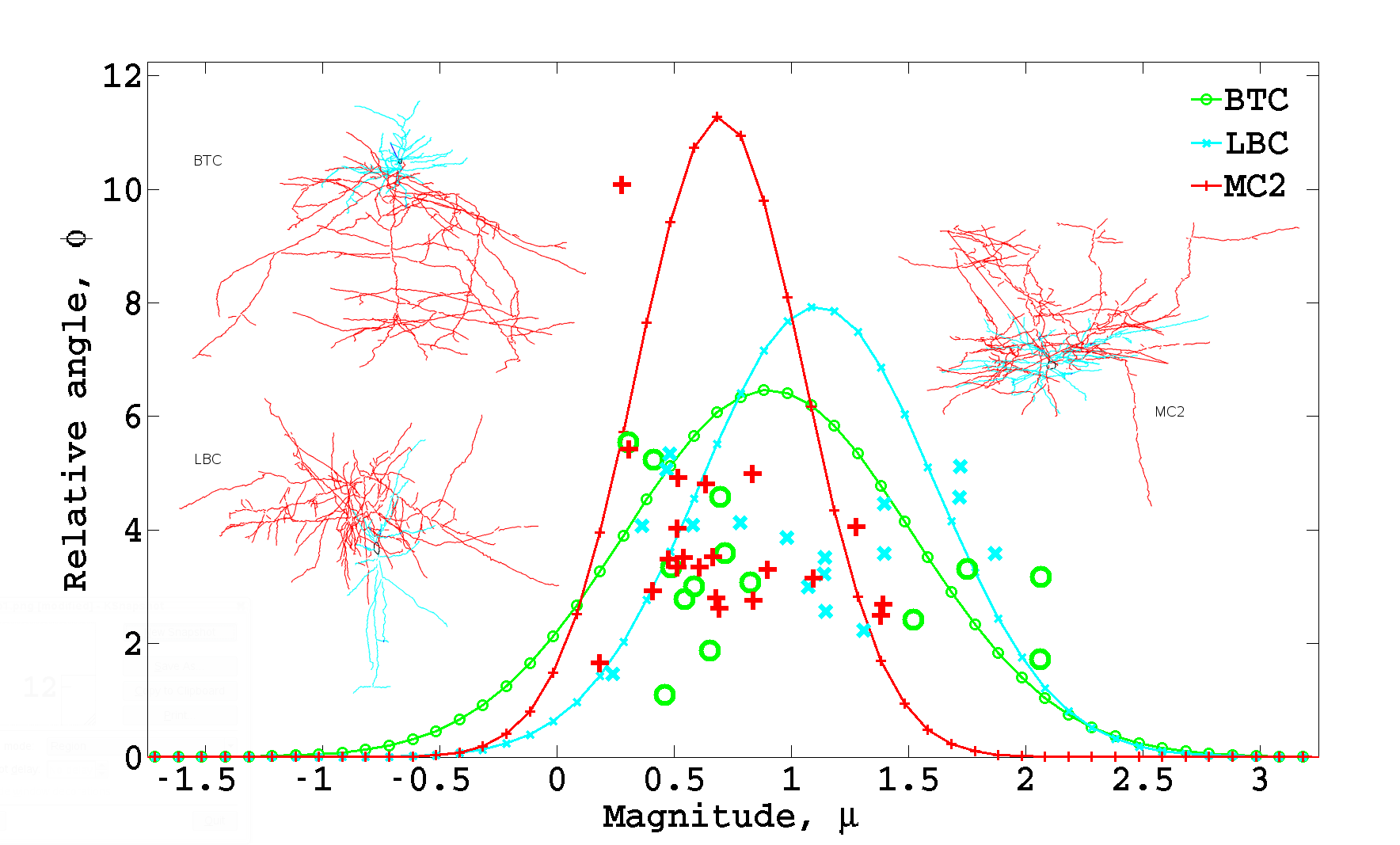}   \caption{The distribution of axonal-dendritic similarity among cells
 in the three neuronal classes.\label{fig:distrib}}
\end{centering}
\end{figure*}

The result of the above procedure is presented in the scatterplot
(i.e. the angle \emph{versus} the magnitude of difference vectors
for each neuron) in the composited Figure~\ref{fig:distrib}, which
also shows the relative densities of the magnitudes of the
difference vectors obtained by using a simple Gaussian fit.  Recall
that total morphologic similarity between the axonal and dendritic
trees for each cell would result in a magnitude of difference vector
equal to zero. The scatterplot in Figure~\ref{fig:distrib} shows
that most cells resulted with magnitude between 0 and 2 and angle
between 0 and 6, with an outlier with angle near to 10 in the case
of the MC2 cell category.  As shown by the respective densities, the
three classes of neuronal cells resulted with distinct dispersions,
with the MC2  class being characterized by the smallest variance
while the BTC class shows the largest variance.  In order to obtain
a more objective indication of the separation between the three
neuronal cell categories, we performed the ANOVA variance test
(e.g.~\cite{Hogg:1987}) and obtained the results ($p-$values) shown
in Table~\ref{table:anova}. These values indicate that the MC2 and
LBC classes are those less likely to have come from the same
population of cells, with a respective $p-$value of only 0.0088.
Substantially higher values were obtained for the other pairs of
categories, suggesting less separability in those cases.

\begin{table}[htb]
\begin{center}
\begin{tabular}{llll}\hline\hline
Class &  BTC &   LBC  &  MC2     \\ \hline
BTC   &      & 0.2676 & 0.1629   \\ \hline
LBC   &      &        & 0.0088   \\ \hline
MC2   &      &        &          \\ \hline
\hline
\end{tabular}
\caption{The resulting $p-$values obtained through ANOVA for the 3 classes of neurons. \label{table:anova}}
\end{center}
\end{table}

Figure~\ref{fig:distrib} also shows an example of neuronal cell for
each of the anatomical classes, lying close to the center of the
respective Gaussian distribution fit.  Despite the difficulty of
comparing the original 3D axonal and dendritic arborizations from such
2D projections, the larger difference between the two arborizations
obtained in the case of the LBC category is still clear, as well as
the greater similarity observed for the MC2 cell example.

The obtained results suggest varying degrees of morphological
relationship between the axonal and dendritic arborizations among all
the considered individual cells and also between the three classes
BTC, LBC and MC2.  In the latter case, the classes LBC e MC2 resulted
as being strongly statistically distinct.  In brief, the methodology
reported in this work has been found to provide a sensitive means for
inferring differences between the axonal and dendritic structures.
This is a particularly important result, as such differences may be
related to important differentiation factors during the neuronal
growth, including those caused by genetic modifications (cell
predestination), environmental constraints (e.g., interactions with
other cells and target specificity), as well as the history of presented
stimuli.  The identification of such differences also present
potential as a complementary subsidy for neuronal cell classification
and diagnosis of abnormalities in neurological diseases, which
correspond to interesting topics for further research.

\begin{acknowledgments}

Luciano da F. Costa is grateful to FAPESP (05/00587-5) and CNPq
(308231/03-1) for financial support.  Marconi S. Barbosa is grateful
to FAPESP (02/02504-1,03/02789-9) for sponsoring his post-doc programme.

\end{acknowledgments}

\bibliographystyle{unsrt} \bibliography{apl}

\end{document}